\documentclass[a4paper,11pt]{article}

\usepackage{amsmath}
\usepackage{amssymb}
\usepackage{color}
\usepackage{cite}
\usepackage{hyperref}

\makeatletter
\@addtoreset{equation}{section}
\renewcommand{\theequation}{\thesection.\@arabic\c@equation}
\makeatother
\makeatletter
\renewcommand\appendix{\par
  \setcounter{section}{0}%
  \setcounter{subsection}{0}%
  \gdef\thesection{Appendix \@Alph\c@section }
  \renewcommand{\theequation}
  {\Alph{section}.\arabic{equation}}
}
\makeatother

\def \be {\begin{equation}}
\def \ee {\end{equation}}
\def \ba {\begin{array}}
\def \ea {\end{array}}
\def \bea{\begin{eqnarray}}
\def \eea{\end{eqnarray}}

\def \b {\beta}
\def \g {\gamma}

\def \d {\delta}

\def \m {\mu}

\def \s {\sigma}

\def \r {\rho}

\def \f {\frac}

\def \nn {\nonumber}

\def \hs {\hspace}

\def \inf {\infty}

\def \Tr {{\textrm{Tr}}}

\def \tr {{\textrm{tr}}}

\title{\textbf{Single Interval R\'enyi Entropy At Low Temperature}}
\author{
Bin Chen$^{1,2,3}$\footnote{bchen01@pku.edu.cn}\,
and
Jie-qiang Wu$^{1}$\footnote{jieqiangwu@pku.edu.cn}
}
\date{}

\begin{document}

\maketitle

\begin{center}
{\it
$^{1}$Department of Physics and State Key Laboratory of Nuclear Physics and Technology, Peking University, 5 Yiheyuan Rd, Beijing 100871, P.R.\! China\\
\vspace{2mm}
$^{2}$Collaborative Innovation Center of Quantum Matter, 5 Yiheyuan Rd, \\Beijing 100871, P.~R.~China\\
$^{3}$Center for High Energy Physics, Peking University, 5 Yiheyuan Rd, \\Beijing 100871, P.~R.~China
}
\vspace{10mm}
\end{center}

\begin{abstract}
In this paper, we calculate the R\'enyi entropy of one single interval on a circle at finite temperature in 2D CFT. In the low temperature limit, we expand the thermal density matrix level by level in the vacuum Verma module, and calculate the first few leading terms in $e^{-\pi/TL}$ explicitly. On the other hand, we compute the same R\'enyi entropy holographically. After considering the dependence of the R\'enyi entropy on the temperature, we manage to fix the interval-independent constant terms in the classical part of holographic R\'enyi entropy. We furthermore extend the analysis in \cite{Dong} to higher orders and find exact agreement between the results from field theory and bulk computations in the large central charge limit. Our work provides another piece of evidence to support holographic computation of R\'enyi entropy in AdS$_3$/CFT$_2$ correspondence, even with thermal effect. 
\end{abstract}

\baselineskip 18pt
\thispagestyle{empty}

\newpage

\section{Introduction}

R\'enyi entropy is an important quantity in the study of the entanglement of a subsystem $A$ with its compliment $B$\cite{nielsen2010quantum,petz2008quantum}. The usual entanglement entropy of $A$ is defined as the Von Neumann entropy of the reduced density matrix
\be
S_A=-\Tr_A \r_A\log \r_A.
\ee
In practice, it is hard to compute the entropy using the above relation. Instead, one
may start from the R\'enyi entropy, which is defined as
\be
S_A^{(n)}=-\f{1}{n-1} \log \Tr_A \r_A^n,
\ee
and take the $n\to 1$ limit to read the entanglement entropy
\be
S_A=\lim_{n \to 1} S_A^{(n)}.
\ee
Besides providing an effective way to compute the entanglement entropy, the R\'enyi entropy is also essential to understand the spectral properties of the reduced density matrix. Once we know the R\'enyi entropy $S_A^{(n)}$ for all $n\in N$, we may find the spectrum of $\r_A$.

Very recently, it has been discovered that the R\'enyi entropy plays an indispensable role in the study of the AdS/CFT correspondence, especially the AdS$_3$/CFT$_2$ correspondence. In \cite{Faulkner,Hartman}, it has been proved that for multiple intervals in two-dimensional(2D) CFT, the leading contributions of their R\'enyi entropies in the large central charge limit are given by the classical action of corresponding gravitational configurations. In 2D CFT, the R\'enyi entropy is usually given by the partition function on a higher genus Riemann surface via the replica trick\cite{Calabrese:2009ez}. From the AdS$_3$/CFT$_2$ correspondence, the bulk gravitational configuration is the classical solution whose asymptotic boundary is exactly the Riemann surface got in CFT\cite{Headrick:2010zt}. The study in \cite{Faulkner,Hartman} actually proved the holographic computation of R\'enyi entropy to leading order in 2D CFT\cite{Ryu:2006bv,Ryu:2006ef}.

In the AdS$_3$/CFT$_2$ correspondence, the central charge of 2D CFT is inversely proportional to the three-dimensional Newton coupling constant
\be
c=\frac{3l}{2G}.
\ee
In the large central charge limit, the gravity is weakly coupled. The 1-loop correction to the holographic R\'enyi entropy could be obtained from the 1-loop partition function around the classical gravitational configuration. In \cite{Headrick:2010zt,Dong}, the 1-loop correction to the holographic R\'enyi entropy of  two intervals has been studied. Such correction presents as the terms independent of the central charge in the CFT computation. In \cite{Zhang}, by using the operator product expansion(OPE) of the twist operators in the small interval limit\cite{Calabrese:2010he}, the R\'enyi entropy in two-interval case has been computed order by order in  small cross ratio. It has been found in \cite{Zhang}, both the leading contribution and the subleading correction are in perfect match with the holographic results. This provides strong support that the holographic computation of quantum correction to the R\'enyi entropy is correct. Further evidences have been obtained from the study of the holographic R\'enyi entropy in the  CFT with $W$ symmetry\cite{Chen:2013dxa,Perlmutter:2013paa},  in the AdS/LCFT correspondence\cite{Chen:2014kja} and in the case with scalar coupling\cite{Beccaria:2014lqa}.

In this work, we shall discuss the R\'enyi entropy of one single interval on a circle in 2D CFT at low temperature. The holographic computation has been worked out in some details in \cite{Dong}. However, there is short of direct CFT computation\footnote{For a concrete computation of R\'enyi entropy of free boson on a torus, see \cite{Datta:2013hba}.}. We address this unsolved issue. We are inspired by the recent study of thermal correction to the entanglement entropy in \cite{Cardyun}. The essential point in the treatment of \cite{Cardyun} is that at low temperature, there are excitations above the vacuum so that the thermal density matrix should be expanded level by level, and the computation boils down to the correlation function of vertex operators corresponding to the excitations. For our purpose, we focus on the vacuum Verma module and consider the excitations to level 4. In the large central charge limit, we read the order ${\mathcal O}(c)$ contribution to order $e^{-8\pi/TL}$, and the $c$-independent contribution to order $e^{-6\pi/TL}$.

On the other hand, we compute the same R\'enyi entropy holographically. Our strategy is to calculate it at high temperature first and then do a modular transformation to read the result at low temperature\footnote{Certainly we can work directly with the  low temperature case directly. It turns out that the two cases are related by a modular transformation, as shown in \cite{Dong}.}.  The classical part is determined by the differential equation
\be
 \frac{\partial S_n^{\gamma}}{\partial z_i}=-\frac{cn}{6(n-1)}\gamma_i
\ee
where $\g_i$ is the accessory parameters. However in the finite temperature case, we notice that this equation alone cannot fix the $y$-independent constant terms, where $2y$ is the distance of the interval. At high temperature, considering the dependence of the R\'enyi entropy on the radius of the circle, we propose the relation
\be \frac{\partial S_n}{\partial R}=\frac{c}{12\pi}\frac{n}{n-1}\beta(\tilde{\delta}-{\tilde \delta}_{n=1}), \ee
where $\tilde{\delta}$ consists of all $z$-independent terms. This equation allows us
to determine all the $y$-independent constant terms. Consequently we can extend the analysis in \cite{Dong} to higher orders. By a modular transformation, we obtain the holographic R\'enyi entropy at low temperature and  find perfect agreement with the results from the field theory computation.

This work is organized as follows. In section 2, we compute the R\'enyi entropy in 2D CFT at low temperature. In section 3, we generalize the analysis in \cite{Dong} to higher orders and compute the holographic R\'enyi entropy. In section 4, we end with conclusion and discussion. In Appendices, we present some technical details.


\section{R\'enyi entropy at low temperature in CFT: single interval case}

In this section, we try to calculate single interval entangle entropy and Renyi entropy on a circle at low temperature. The circle length is $L$, the interval length is $l$, the temperature is $T=\frac{1}{\b}$. In the field theory, we can compute the  Renyi entropy and entanglement entropy by using the replica trick as in \cite{Cardy,Cardyun}.  By definition the thermal density matrix is
\be \rho=\frac{e^{-\b H}}{\Tr e^{-\beta H}}=\frac{1}{\Tr e^{-\beta H}}\sum\mid\phi\rangle\langle\phi\mid e^{-\beta E_{\phi}} \ee
where the summation is over all the excitations in the theory.
On a cylinder the energy spectrum is read by
\be H=\frac{2\pi}{L}(L_0+\widetilde{L_0}-\frac{c}{12}) \ee
In this work, we only focus on the excitations from the vacuum Verma module, so
the thermal density matrix is
\bea\label{entropy} \rho=\frac{\mid 0\rangle\langle 0\mid+\sum|\phi_2\rangle\langle \phi_2|e^{\frac{-4\pi\beta}{L}}+\sum|\phi_3\rangle\langle \phi_3|e^{\frac{-6\pi\beta}{L}}+\sum|\phi_4\rangle\langle \phi_4|e^{\frac{-8\pi\beta}{L}}+O(e^{\frac{-10\pi\beta}{L}})}
{(1+2e^{\frac{-4\pi\beta}{L}}+2e^{\frac{-6\pi\beta}{L}}+5e^{\frac{-8\pi\beta}{L}}+O(e^{\frac{-10\pi\beta}{L}}))}
\eea
where $|\phi_i\rangle$ stands for all normalized states with the total level $i$. More explicitly, the level two states include
\bea &~&\mid 2\rangle=\sqrt{\frac{2}{c}}L_{-2}\mid 0\rangle,~\mid \widetilde{2}\rangle=\sqrt{\frac{2}{c}}\widetilde{L_{-2}}\mid 0\rangle, \notag\eea
the level three states include
\bea
 &~& \mid 3\rangle=\sqrt{\frac{1}{2c}}L_{-3}\mid 0\rangle,~\mid \widetilde{3}\rangle=\sqrt{\frac{1}{2c}}\widetilde{L_{-3}}\mid 0\rangle, \notag \eea
 and the level four states include
 \bea
&~& \mid 4,1\rangle=\sqrt{\frac{1}{5c}}L_{-4}\mid 0\rangle,~\mid\widetilde{4,1}\rangle=\sqrt{\frac{1}{5c}}\widetilde{L_{-4}}\mid 0\rangle, \notag \\
&~& \mid 4,2\rangle=(\frac{c^2}{2}+\frac{11}{5}c)^{-\frac{1}{2}}(L_{-2}L_{-2}-\frac{3}{5}L_{-4}) \mid 0\rangle, \notag \\
&~& \mid \widetilde{4,2}\rangle=(\frac{c^2}{2}+\frac{11}{5}c)^{-\frac{1}{2}}(\widetilde{L_{-2}}\widetilde{L_{-2}}-\frac{3}{5}\widetilde{L_{-4}}) \mid 0\rangle, \notag \\
&~& \mid 2,\widetilde{2}\rangle=\frac{2}{c}L_{-2}\widetilde{L_{-2}} {\mid}0\rangle.  \eea

We divide the space into two parts: A  and its complement B. The reduced density matrix for A is
\be \rho_A=\tr_B\rho \ee
and the Renyi entropy is defined as
\be\label{Renyi} S_n=\frac{1}{1-n}\log\Tr({\rho_A}^n). \ee
In the thermal case, we find that
\bea\label{logtrace} \log\Tr(\rho_A)^n
=\log\Tr(\tr_B\mid 0\rangle\langle0\mid)^n+A_2e^{-\frac{4\pi\beta}{L}}+A_3e^{-\frac{6\pi\beta}{L}}+A_4e^{-\frac{8\pi\beta}{L}}
+O(e^{-\frac{10\pi\beta}{L}}) \eea
where the first term is the zero temperature contribution from the vacuum, and the other terms are the thermal corrections at low temperature. Here
\bea A_2&=&n\frac{\sum \Tr[\tr_B\mid{\phi_2}\rangle\langle\phi_2\mid(\tr_B\mid 0\rangle\langle0\mid)^{n-1}]}{\Tr(\tr_B\mid 0\rangle\langle 0\mid)^n} -2n, \eea
\bea A_3&=&n\frac{\sum \Tr[\tr_B\mid{\phi_3}\rangle\langle\phi_3\mid(\tr_B\mid 0\rangle\langle 0\mid)^{n-1}]}{\Tr(\tr_B\mid 0\rangle\langle 0\mid)^n} -2n, \eea
and
\bea  A_4&=&-3n+\big(n\frac{\Tr[\tr_B\mid{4,1}\rangle\langle 4,1\mid(\tr_B\mid 0\rangle\langle 0\mid)^{n-1}+\tr_B\mid{4,2}\rangle\langle 4,2\mid(\tr_B\mid 0\rangle\langle 0\mid)^{n-1}]}{\Tr(\tr_B\mid 0\rangle\langle 0\mid)^n} \notag \\
&~&+\frac{n}{2}\sum_{j=1}^{n-1}\frac{\Tr[\tr_B{\mid}2\rangle\langle  2\mid(\tr_B{\mid}0\rangle\langle 0\mid)^{j-1}
\tr_B{\mid}2\rangle\langle 2\mid(\tr_B{\mid}0\rangle\langle   0\mid)^{n-1-j}]}{\Tr[\tr_B{\mid}0\rangle\langle 0\mid]^n} \notag \\
&~&-\frac{1}{2}n^2\frac{\Tr[\tr_B\mid{2}\rangle\langle 2\mid(\tr_B\mid 0\rangle\langle 0\mid)^{n-1}]}{\Tr(\tr_B\mid 0\rangle\langle 0\mid)^n}+\mbox{anti-holomorphic terms}\big). \eea
In the above, we have used the fact the computation could be decomposed into holomorphic and anti-holomorphic sectors. In $A_i$'s, there are two kinds of multi-point functions. One is the two-point function on the $n$-sheet Riemann surface
with two vertex operators inserting at the infinity future and infinity past on one sheet
\be \frac{\tr[\tr_B\mid O_1\rangle\langle O_2\mid(\tr_B\mid 0\rangle\langle0\mid)^{(n-1)}]}{\tr(\tr_B\mid 0\rangle\langle 0\mid)^n}. \ee
The other is the four-point function
\be \frac{\tr[\tr_B\mid O_1\rangle\langle  O_2\mid(\tr_B\mid 0\rangle\langle 0\mid)^{(j-1)}\tr_B\mid \widetilde{O_1}\rangle\langle \widetilde{O_2}\mid(\tr_B\mid 0\rangle\langle 0\mid)^{(n-j-1)}]}{\tr(\tr_B\mid 0\rangle\langle  0\mid)^n} \ee
where $O_1, O_2$ are two vertex operators inserting at the infinity future and the infinity past on one sheet, while $ \widetilde{O_1}, \widetilde{O_2}$ are two other vertex operators inserting at the infinity future and the infinite past on another sheet.

 For convenience,  let us make a conformal transformation
\be u=e^{-i\frac{2\pi}{L}z} \ee
In this coordinate, the two branch points are $e^{i\theta_1}$ and $e^{i\theta_2}$ with
\be \theta_2-\theta_1=\frac{2\pi l}{L}, \ee
and the vertex operators are now inserted at the origin and the infinity respectively.
In the Appendices, we discuss how to read the explicit form of the vertex operators in this new coordinate. They
 can be written as (\ref{vertex0}, \ref{vertexinf1}, \ref{vertexinf2})
\bea\label{vertex} L_{-2}\mid 0\rangle&\rightarrow&T(u)\mid_{u=0}, \notag \\
&~&w^4T(w)\mid_{w\rightarrow\inf}, \notag \\
L_{-3}\mid 0\rangle&\rightarrow&\partial T(u)\mid_{u=0}, \notag \\
&~&-w^6\partial T(w)-4w^5T(w)\mid_{w\rightarrow\inf}, \notag \\
L_{-4}\mid 0\rangle&\rightarrow&\frac{1}{2}\partial^2T(u)\mid_{u=0}, \notag \\
&~&\frac{1}{2}w^8\partial^2T(w)+5w^7\partial T(w)+10w^6T(w)\mid_{w\rightarrow\inf}, \notag \\
L_{-2}L_{-2}\mid 0\rangle& \rightarrow& :T(u)^2:\mid_{u=0}, \notag \\
&~&w^8:T(w)^2:+3w^7\partial T(w)+6w^6T(w)\mid_{w\rightarrow\inf}, \notag \\
L_{-2}L_{-2}-\frac{3}{5}L_{-4}\mid 0\rangle&\rightarrow& :T(u)^2:-\frac{3}{10}\partial^2 T(u)\mid_{u=0}, \notag \\
&~&w^8(:T(w)^2:-\frac{3}{10}\partial^2T(w))\mid_{w\rightarrow\inf}. \eea

We can further make a coordinate transformation, taking the surface into one complex plane
\be \zeta=(\frac{u-e^{i\theta_2}}{u-e^{i\theta_1}})^{\frac{1}{n}}. \ee
 Therefore we may compute the multi-point functions in full complex plane, and then make a inverse conformal transformation  to read the functions on the $n$-sheet Riemann surface.
For example, in the full complex plane
\be \langle T(\zeta_2)T(\zeta_1)\rangle=\frac{c}{2}\frac{1}{(\zeta_2-\zeta_1)^4}. \ee
Under a conformal transformation
\bea T(u)&=&T(\zeta)(\frac{\partial \zeta}{\partial u})^2+\frac{c}{12}\{\zeta,u\} \notag \\
&=&T(\zeta)(\frac{\partial \zeta}{\partial u})^2+\frac{c}{24}\frac{(e^{i\theta_2}-e^{i\theta_1})^2}{(u-e^{i\theta_2})^2(u-e^{i\theta_1})^2}. \eea
Since $\langle T(\zeta)\rangle=0$, then we get
\bea \label{twop} &~&\langle T(w)T(u)\rangle=(\frac{c}{24})^2\frac{((e^{i\theta_2}-e^{i\theta_1})^2)^2}{(w-e^{i\theta_2})^2(w-e^{i\theta_1})^2(u-e^{i\theta_2})^2(u-e^{i\theta_1})^2} \notag \\
 &~&+(\frac{1}{n}(e^{i\theta_2}-e^{i\theta_1}))^4\langle T(\zeta_2)T(\zeta_1)\rangle \notag \\
&~&\times(u-e^{i\theta_2})^{2(\frac{1}{n}-1)}(u-e^{i\theta_1})^{-2(\frac{1}{n}+1)}
(w-e^{i\theta_2})^{2(\frac{1}{n}-1)}(w-e^{i\theta_1})^{-2(\frac{1}{n}+1)}.
\eea

For  $A_2$, its holomorphic part is a two-point function on $n$-sheet Riemann surface
\bea\label{second} &~&\frac{\tr[\tr_B\mid{2}\rangle\langle 2\mid(\tr_B\mid 0\rangle\langle 0\mid)^{n-1}]}{\Tr(\tr_B\mid 0\rangle\langle 0\mid)^n}\notag \\
& =&\lim_{w\rightarrow\inf,u=0}\frac{2}{c}\langle w^4T(w)T(u)\rangle
_{n-sheet} \notag \\
&=&\frac{c}{18}(1-\frac{1}{n^2})^2\sin^4(\frac{\pi l}{L})
+\frac{1}{n^4}\frac{\sin^4(\frac{\pi l}{L})}{\sin^4(\frac{\pi l}{nL})}
\eea

For $A_3$, we compute it in a similar way and get
\bea &~&\frac{\Tr[\tr_B\mid{3}\rangle\langle 3\mid(\tr_B\mid 0\rangle\langle 0\mid)^{n-1}]}{\Tr(\tr_B\mid 0\rangle\langle 0\mid)^n} \notag \\
&=&\frac{1}{2c}\lim_{w\rightarrow \inf u=0}\langle (-w^6\partial T(w)-4w^5T(w))\partial T(u)\rangle \notag \\
&=& \frac{2}{9}c(1-\frac{1}{n^2})^2\sin^4(\frac{\pi l}{L})\cos^2(\frac{\pi l}{L})
+\frac{1}{n^6}\big(5\frac{\sin^6(\frac{\pi l}{L})}{\sin^6(\frac{\pi l}{nL})}
-4\frac{\sin^6(\frac{\pi l}{L})}{\sin^4(\frac{\pi l}{nL})}\big) \notag \\
&~&-\frac{8}{n^5}\frac{\sin^5(\frac{\pi l}{L})}{\sin^5(\frac{\pi l}{nL})}\cos(\frac{\pi l}{L})\cos(\frac{\pi l}{nL}) +\frac{4}{n^4}\frac{\sin^4(\frac{\pi l}{L})}{\sin^4(\frac{\pi l}{nL})}\cos^2(\frac{\pi l}{L}).
\eea

For $A_4$, the computation becomes quite complicated and we are satisfied to get the leading $c$ term. Let us calculate the terms in $A_4$ one by one. First of all
\bea &~&\frac{\Tr[\tr_B\mid{4,1}\rangle\langle 4,1\mid(\tr_B\mid 0\rangle\langle 0\mid)^{n-1}]}{\Tr(\tr_B\mid 0\rangle\langle 0\mid)^n} \notag \\
&=&\frac{1}{5c}\langle (\frac{1}{2}w^8\partial^2T(w)+5w^7\partial T(w)+10w^6T(w))\frac{1}{2}\partial^2T(u)\rangle\mid_{w\rightarrow\inf,u=0} \notag \\
&=&\frac{c}{45}(1-\frac{1}{n^2})^2\sin^4(\frac{\pi l}{L})(3\cos(\frac{2\pi l}{L})+2)^2+O(c^0) \eea
Secondly
\bea &~&\frac{\Tr[\tr_B\mid{4,2}\rangle\langle 4,2\mid(\tr_B\mid 0\rangle\langle 0\mid)^{n-1}]}{\Tr(\tr_B\mid 0\rangle\langle 0\mid)^n} \notag \\
&=&(\frac{1}{2}c^2+\frac{11}{5})^{-1}\langle w^8(:T(w)^2:-\frac{3}{10}\partial^2T(w))(:T(u)^2:-\frac{3}{10}\partial^2T(u))\rangle
\mid_{w\rightarrow\inf,u=0} \notag \\
&=&\frac{c^2}{648}(1-\frac{1}{n^2})^4\sin^8\frac{\pi l}{L}
+\frac{11c}{1620}(1-\frac{1}{n^2})^4\sin^8\frac{\pi l}{L}
+\frac{c}{9}\frac{1}{n^4}(1-\frac{1}{n^2})^2\frac{\sin^8\frac{\pi l}{L}}{\sin^4\frac{\pi l}{nL}} \notag \\
&~&+O(c^0) \eea
In the calculation, we have used Eqs. (\ref{twopoint1}) (\ref{twopoint2}) and (\ref{summation1}) (\ref{summation2}).
Next
\bea &~&\sum_{j=1}^{n-1}\frac{\Tr[\tr_B{\mid}2\rangle\langle 2\mid(\tr_B{\mid}0\rangle\langle 0\mid)^{j-1}
\tr_B{\mid}2\rangle\langle 2\mid(\tr_B{\mid}0\rangle\langle 0\mid)^{n-1-j}]}{\Tr[\tr_B{\mid}0\rangle\langle 0\mid]^n} \notag \\
&=&\sum_{j=1}^{n-1}(\frac{2}{c})^2\langle w^4T^{(1)}(w)T^{(1)}(u)\tilde{w}^4T^{(j)}(\tilde{w})T^{(j)}(\tilde{u})\rangle \notag \\
&=&\frac{c^2}{324}(n-1)(1-\frac{1}{n^2})^4\sin^8\frac{\pi l}{L} +\frac{c}{9}(n-2)(1-\frac{1}{n^2})^2\frac{1}{n^4}
\frac{\sin^8\frac{\pi l}{L}}{\sin^4\frac{\pi l}{nL}} \notag \\
&~&+\frac{c}{405}(1-\frac{1}{n^2})^2\frac{(n^2+11)(n^2-1)}{n^4}\sin^8\frac{\pi l}{L} \notag \\
&~&+\frac{c}{27}(1-\frac{1}{n^2})^2\sin^4\frac{\pi l}{L}
[(1-\frac{1}{n^2})\cos(\frac{2\pi l}{L})+(2+\frac{1}{n^2})]+O(c^0)
\eea

The antiholomophic terms give the similar contributions. Taking all the contributions into account, we obtain the R\'enyi entropy, which could be classified into the tree-level part and the 1-loop part
\bea &~&S_n=\frac{1}{1-n}\log~tr\rho_A^n = S_n^{\mbox{\tiny tree}}+ S_n^{\mbox{\tiny 1-loop}}\notag \eea
The tree-level part is proportional to the central charge
\bea
S_n^{\mbox{\tiny tree}}&=&\frac{c(1+n)}{n}\left\{\frac{1}{12}\log\sin^2\frac{\pi l}{L}+const\right.\notag \\
&~&-\frac{1}{9}\frac{(n^2-1)}{n^2}\left(\sin^4(\frac{\pi l}{L})
e^{-\frac{4\pi\beta}{L}}
+4\sin^4(\frac{\pi l}{L})\cos^2(\frac{\pi l}{L})
e^{-\frac{6\pi\beta}{L}} \right.\notag \\
&~&+\left(\frac{-11-2n^2+1309n^4}{11520n^4}\cos(\frac{8\pi l}{L})-\frac{-11+28n^2+119n^4}{1440n^4}\cos(\frac{6\pi l}{L}) \right.\notag\\
&~&-\frac{77-346n^2+197n^4}{2880n^4}\cos(\frac{4\pi l}{L})-\frac{-77+436n^2+433n^4}{1440n^4}\cos(\frac{2\pi l}{L}) \notag \\
&~&\left. \left. \left. +\frac{-77+466n^2+907n^4}{2304n^4}\right)e^{-\frac{8\pi\beta}{L}}\right)+O(e^{-\frac{10\pi\beta}{L}})\right\}.
\label{tree}
\eea
And the 1-loop part is independent of $c$
\bea
S_n^{\mbox{\tiny 1-loop}}&=&-\frac{2n}{n-1}\left(\frac{1}{n^4}\frac{\sin^4(\frac{\pi l}{L})}{\sin^4\frac{\pi l}{nL}}-1\right)
e^{-\frac{4\pi\beta}{L}} \notag \\
&~&-\frac{2n}{n-1}\left(\frac{4}{n^4}(\frac{\sin\frac{\pi l}{L}}{\sin\frac{\pi l}{nL}})^4
\cos^2\frac{\pi l}{L}
-\frac{8}{n^5}(\frac{\sin\frac{\pi l}{L}}{\sin\frac{\pi l}{nL}})^5
\cos\frac{\pi l}{nL}\cos\frac{\pi l}{L} \right.\notag \\
&~&\left. -\frac{4}{n^6}(\frac{\sin\frac{\pi l}{L}}{\sin\frac{\pi l}{nL}})^4
\sin^2\frac{\pi l}{L}
+\frac{5}{n^6}(\frac{\sin\frac{\pi l}{L}}{\sin\frac{\pi l}{nL}})^6-1\right)
e^{-\frac{6\pi\beta}{L}}
+O(e^{-\frac{8\pi\beta}{L}}). \label{1loop}
 \eea
As shown in \cite{Cardyun}, there is a symmetry $l \to nL-l$ in the $n$-th R\'enyi entropy.

The entanglement entropy could be read easily
\bea S_{EE}&=&\lim_{n\rightarrow 1}S_n \notag \\
&=&c\left(\frac{1}{6}\log~\sin^2\frac{\pi l}{L}+const\right)\nonumber\\
& &+8(1-\frac{\pi l}{L}\cot(\frac{\pi l}{L}))e^{-\frac{4\pi\beta}{L}}
+12(1-\frac{\pi l}{L}\cot(\frac{\pi l}{L}))e^{-\frac{6\pi\beta}{L}}+O(e^{-\frac{8\pi\beta}{L}}). \label{EE}
 \eea
Due to the thermal correction terms, the symmetry $l \to L-l$ is broken. Such correction terms are independent of the central charge, and are expected to be captured by the quantum correction in the holographic computation. Different from entanglement entropy, there appear thermal correction terms even at the leading order in the R\'enyi entropy.

So far, we have calculated the low temperature R\'enyi and entangle entropy  to order $e^{-\frac{8\pi\beta}{L}}$ for $O(c)$ term, and to order $e^{-\frac{6\pi\beta}{L}}$ for $O(1)$ term. The Renyi entropy is slightly different from Cardy and Herzog's result at the leading order of the thermal correction, because we are now considering the energy-momentum tensor which is not a primary field. But the entanglement entropy is the same as Cardy and Herzog's universal result\cite{Cardyun}. In the next section, we compute these quantities holographically in the bulk, and find exact agreement.

\section{Holographic computation}

The Renyi entropy in 2D CFT can be calculated holographically by using AdS/CFT correspondence in the large $c$ limit\cite{Faulkner,Dong}. For a fixed Riemann surface at the boundary originating from the replica trick, one  needs to find the dominant gravitational configuration, whose gravity action gives us the classical R\'enyi entropy. Moreover,  the one-loop partition function of various fluctuations around the configuration leads to the 1-loop correction to the holographic R\'enyi entropy. For the classical contribution, the holographic R\'enyi entropy is given by the following relation
\be\label{branch} \frac{\partial S_n^{\gamma}}{\partial z_i}=-\frac{cn}{6(n-1)}\gamma_i \ee
where $\g_i$'s are called the accessory parameters appearing in the equation
\be\label{diff} \psi^{''}(z)+\frac{1}{2}T(z)\psi(z)=0, \ee
\be T(z)=\sum_i(\frac{\Delta}{(z-z_i)^2}+\frac{\gamma_i}{z-z_i}) \ee
in the zero temperature case\cite{Faulkner} with $\Delta=\frac{1}{2}(1-\frac{1}{n^2})$, and
\be T(z)=\sum_i(\Delta \wp(z-z_i)+\gamma_i\zeta(z-z_i)+\delta) \ee
in the finite temperature case, with $\wp,~\zeta$ being the Weierstrass elliptic function and Weierstrass zeta function respectively and $\d$ being an additional constant\cite{Dong}.

 Choosing two solutions $\psi_1$ and $\psi_2$ of the equation (\ref{diff}), locally they define a conformal transformation $w=\frac{\psi_1}{\psi_2}$. The conformal transformation defines a global Schottky uniformization, which respects the replica symmetry. Holographically, every Schottky uniformization can be extended into the bulk \cite{Krasnov}. For one boundary configuration there can be different Schottky uniformizations and correspondingly different  bulk configurations for the same boundary configuration. With proper regulation the classical gravity action reduces to the Liouville action\cite{Liou}, which leads to the relation (\ref{class}). The $\g_i$'s are determined by imposing the monodromy condition. For two interval at zero temperature case, it has been discussed carefully in \cite{Faulkner}.

In the case of one finite circle at finite temperature, there are two cycles on each sheet
\be u\sim u+mR+in\beta, \ee
with $R$ being the length of the circle and $\beta$ being the imaginary time. We can set trivial monodromy along one cycle so that the other cycle is  the generator of Schottky group. For example, for single interval finite temperature cylinder, the Riemann surface corresponding to the $n$-th R\'enyi entropy is of genus $n$, and the Schottky generators are the non-trivial cycles in $n$ sheets.
At high temperature, the time direction is of trivial monodromy, and at low temperature, the spatial direction is of trivial monodromy. This is because at high temperature above the Hawking-Page transition, the bulk spacetime is actually a black hole, while at low temperature, it is the thermal AdS spacetime. However, for the finite temperature case the equation (\ref{branch}) is not enough to fix the Renyi entropy completely. There are interval-independent terms in the classical action, which appear as the integration constants and cannot be fixed by the equation (\ref{branch}). In this work, we propose one more relation to  determine all the terms uniquely.

For 1-loop quantum correction, after finding all primitive words in Schottky group, then the 1-loop partition function is\cite{Yin,Giombi:2008vd,Dong}
\be\label{1loopY} \log Z\mid_{1-loop}=-\sum_{\gamma}\sum_{m=2}^{\inf}\log\mid 1-q_{\gamma}^m\mid, \ee
where $\gamma$'s are the primitive words, and $q_{\gamma}^{-\frac{1}{2}}$ is the larger eigenvalue of $\gamma$.
There are infinite terms in the summation. However, in the cases of single interval in a circle at low temperature or high temperature, and double intervals with small   cross ratio $x$ at zero temperature, there are only finite terms contributing in each order of  $e^{-\frac{\pi\beta}{R}}$, $e^{-\frac{\pi R}{\beta}}$ or $x$\cite{Dong}.

Furthermore, we need to point out that the holographic R\'enyi entropies at the low temperature and high temperature are dual to each other by the following transformation
\be\label{Sdual} R\rightarrow i\beta, \hs{3ex}~\beta\rightarrow iR. \ee
The classical part and the 1-loop correction of holographic R\'enyi entropy have been computed in \cite{Dong} to order $e^{-\frac{6\pi \b}{L}}$ and $e^{-\frac{4\pi \b}{L}}$   respectively. We will extend their analysis to higher order. By using  the above transformation we can read the entropy at low temperature. However, there is no difficulty to do the calculation at low temperature directly.

\subsection{Classical contribution at high temperature}

In this subsection, we compute the high temperature Renyi entropy of a single interval on a circle, following the strategy and convention in \cite{Dong}. The temperature is T, the length of the circle is R. The interval length is 2y, with two branch points being at $\pm y$. At high temperature, the monodromy along Euclidean time direction should be trivial. With the coordinates
\be u=e^{-2\pi Tz}, ~ u_y=e^{-2\pi Ty}, ~ u_R=e^{-2\pi TR}, \ee
the wave function is
\be \psi_{\pm}=\frac{1}{\sqrt{u}}(u-u_y)^{\Delta_{\m}}(u-\frac{1}{y_y})^{\Delta_{\mp}}\sum_{m=-\inf}^{\inf}\psi_{\pm}^{(m)}(u_y,u_R)u^m. \ee
For convenience, we rewrite the energy momentum tensor as
\bea\label{Tnew} T(z)&=&\frac{1}{2}(1-\frac{1}{n^2})\left(\sum_{m=-\inf}^{+\inf}\frac{4\pi^2 T^2}{uu_yu_R^m+\frac{1}{uu_yu_R^m}-2}
+\sum_{-\inf}^{+\inf}\frac{4\pi^2T^2}{\frac{uu_R^m}{u_y}+\frac{u_y}{uu_R^m}-2}\right) \notag \\
&~&\gamma\left(\sum_{m=-\inf}^{+\inf}\pi T\frac{1+uu_yu_R^m}{1-uu_yu_R^m}
-\sum_{m=-\inf}^{+\inf}\pi T\frac{1+\frac{uu_R^m}{u_y}}{1-\frac{uu_R^m}{u_y}}\right)+\tilde{\delta}. \eea
Here we have extract all of the constant terms independent of $u$ in $\wp$ and $\zeta$ and put them into a new constant $\tilde{\delta}$. The constant $\tilde{\delta}$ consists of an infinite summation with respect to $u_R$. Even though each of the summation in (\ref{Tnew}) including the one in $\tilde{\delta}$ is divergent, $T(z)$ itself is finite. The advantage for this decomposition is that when we integrate $T(z)$ over $z$, only the terms in $\tilde{\delta}$ contribute. We will use this property in the later discussion.

The infinite summation should be convergent at $u_y$ and $\frac{1}{u_y}$. This leads to
\bea \gamma&=&\pi T\left\{-\frac{1-n^2}{n^2}\frac{\cosh({2\pi}Ty)}{\sinh({2\pi}Ty)}
+\frac{1}{6}\frac{1}{n^4}(n^2-1)^2[-2\sinh({8\pi}Ty)+4\sinh({4\pi}Ty)]e^{-{4\pi}TR} \right.\notag \\
&~&+\frac{1}{12}\frac{1}{n^4}(n^2-1)^2[-6\sinh({12\pi}Ty)+8\sinh(8\pi Ty)+2\sinh(4\pi Ty)]e^{-6\pi TR} \notag \\
&~&+\frac{1}{4320}\frac{1}{n^8}(n^2-1)^2[(11+2n^2-1309n^4)2\sinh(16\pi Ty)+6(-11+28n^2+199n^4)2\sinh(12\pi Ty) \notag \\
&~&\left. +2(77-346n^2+197n^4)2\sinh(8\pi Ty)+2(-77+436n^2+433n^4)2\sinh(4\pi Ty)]e^{-8\pi TR}\right\} \notag \\
&~&+O(e^{-10\pi TR}),
 \eea

\bea \label{td}
\tilde{\delta}&=&8\pi^2 T^2\left\{\frac{2}{3}(1-\frac{1}{n^2})^2\cosh^4(2\pi Ty)e^{-4\pi TR}+4(1-\frac{1}{n^2})^2 \sinh^4(2\pi Ty)\cosh^2(2\pi Ty)e^{-6\pi TR} \right.\notag \\
&~&+(1-\frac{1}{n^2})^2\left(\frac{-77+466n^2+907n^4}{1728n^4}+\frac{-11-2n^2+1309n^4}{8640n^4}\cosh(16\pi Ty)\right. \notag \\
&~&+\frac{11-28n^2-199n^4}{1080n^4}\cosh(12\pi Ty)+\frac{-77+346n^2-197n^4}{2160n^4}\cosh(8\pi Ty) \notag\\
&~&\left.\left.+\frac{77-436n^2-433n^4}{1080n^4}\cosh(4\pi Ty)\right)e^{-8\pi TR}+O(e^{-10\pi TR})\right\}-2\pi^2T^2
+O(e^{-10\pi TR}). \nn\\  \eea

As we pointed out before, the differential equation (\ref{branch}) is not enough to determine all the terms uniquely. There exist $y$-independent terms appearing as the integration constants. Here we propose a new differential equation to fix all the terms
\be\label{size} \frac{\partial S_n}{\partial R}=\frac{c}{12\pi}\frac{n}{n-1}\beta(\tilde{\delta}-{\tilde \delta}_{n=1}). \ee
We will discuss how to get this equation in the next subsection. With the equations (\ref{branch}) and (\ref{size}), we can obtain the holographic Renyi entropy
\bea\label{class} &~&S_n\mid_{classical} \notag \\
&=&\frac{c}{6}\frac{1+n}{n}\log\sinh(2\pi TY)+\mbox{const.}-\frac{c}{9}\frac{(n+1)(n^2-1)}{n^3}\left\{\sinh^4(2\pi Ty)e^{-4\pi TR} \right.\notag\\
&~&+4\sinh^4(2\pi Ty)\cosh^2(2\pi Ty)e^{-6\pi TR}+\left(\frac{-11-2n^2+1309n^4}{11520n^4}\cosh(16\pi Ty) \right.\notag \\
&~&-\frac{-11+28n^2+199n^4}{1440n^4}\cosh(12\pi Ty)-\frac{77-346n^2+197n^4}{2880n^4}\cosh(8\pi Ty) \notag \\
&~&\left.\left.-\frac{-77+436n^2+433n^4}{1440n^4}\cosh(4\pi Ty)+\frac{-77+466n^2+907n^4}{2304n^4}\right)e^{-8\pi TR} \right\} \notag \\
&~&+O(e^{-10\pi TR}). \eea

\subsection{Size dependence of Renyi entropy}

In this subsection, we want to give a brief derivation for the equation (\ref{size}). Generally, when we calculate the Renyi entropy, we are computing the partition function on a Riemann surface. The branch cuts' positions, the temperature and the circle length decide the moduli of the Riemann surface. The equation (\ref{branch}) encodes the  dependence of the classical action on the length of the interval. In the finite temperature case, in the same spirit, we should consider the dependence of the classical action on the temperature and the circle length.

For a field theory on a curved spacetime, when we take a variation with respect to the metric, the action changes as
\be \d S_E=-\frac{1}{4\pi}\int d^2\sigma\sqrt{g}\delta g^{ab}T_{ab}^{CFT}. \ee
The $T^{CFT}$ here equal to the $T$ in (\ref{diff}) times $\frac{c}{12}$. This can be seen by comparing the conformal transformation of $T(z)$ \cite{Faulkner} with the one of  $T^{CFT}$.
Now let us take a variation on $g_{11}$ for high temperature case. It changes the length of spatial cycle
\be \delta g_{11}=\frac{2\d R}{R}. \ee
Considering the integration $\int d^2\s \sqrt{g} T(z)$, we notice that
 the first four summations in \ref{Tnew}, after being integrated,  can be written formally as
\be \sum_{m=-\inf}^{+\inf}(f(m+\frac{1}{2})-f(m-\frac{1}{2})) \ee
and cancel each other. Therefore only $\tilde{\delta}$ contribute to the integration
\be \int d^2\s \sqrt{g} T(z)=\beta R\tilde{\delta}. \ee
 Finally
we have
\be dS_E=\frac{c}{12\pi}\beta\tilde{\delta}dR. \ee
By definition
\bea dS_n&=&-\frac{1}{n-1}d(\log Z_n-n\log Z_1)=\frac{1}{n-1}(ndS_E-ndS_E^{(n=1)}). \eea
This leads to the equation (\ref{size})
\be \frac{\partial S_E}{\partial R}=\frac{c}{12\pi}\frac{n}{n-1}\beta(\tilde{\delta}-{\tilde \delta}_{n=1}). \ee

In the above derivation, we only used  the variation with respect to $\delta g_{11}$ but not $\delta g_{22}$. This is because of the monodromy condition. In high temperature the trivial monodromy is along the imaginary time direction, and  in $u$ coordinate  the imaginary time direction is along the angle, so we cannot change the parameter in that direction. On the contrary, in the low temperature case we are only allowed to do variation with respect to $g_{22}$. In this case, we would get the temperature dependence of holographic R\'enyi entropy
\be\label{tem} \frac{\partial S_n}{\partial \b}=-\frac{c}{12\pi}\frac{n}{n-1}R(\tilde{\delta}-{\tilde \delta}_{n=1}) \ee
where to get $\tilde \d$ we should make transformation (\ref{Sdual}) in (\ref{td}).


\subsection{1-loop correction at high temperature}
In this subsection, we calculate the 1-loop correction to holographic R\'enyi entropy at  high temperature.
 In this case the monodromy along the $\beta$ cycle is trivial, and the monodromy along the spatial cycle is
\be L_1=\left(
\begin{array}{ccc}
 (L_1)_{11} & (L_1)_{12} \\
 (L_1)_{21} & (L_1)_{22}
\end{array}
\right)
\ee
where
\bea (L_1)_{11}&=& \frac{nu_y^{1-\frac{1}{n}}}{(1-u_y^2)} \left\{1-\frac{[(n+1)u_y^2+n-1]^2}{4n^2u_y^2}u_R+O(u_R^2)\right\}, \notag \\
 (L_1)_{12}&=&\frac{nu_y}{(1-u_y^2)\sqrt{u_R}}\left\{1-\frac{[(n+1)u_y^2+(n-1)][(n-1)u_y^2+n+1]}{4n^2u_y^2}u_R+O(u_R^2)\right\}, \notag \\
 (L_1)_{21}&=&-(L_1)_{12}, \notag \\
 (L_1)_{22}&=&(L_1)_{11}\mid_{n\rightarrow -n}. \eea

Other Schottky generators can be obtained by a series of movement: first wind  $j$ loops around a branch point, and go along spatial cycle, finally wind back $j$ loops in opposite direction around the same branch point. In other words, they are
\be L_j=M^{j-1}L_1M^{-(j-1)}= \left(
\begin{array}{ccc}
 e^{2\pi i\frac{1}{2}(1+\frac{1}{n})} & 0 \\
 0 & e^{2\pi i\frac{1}{2}(1-\frac{1}{n})}
\end{array}
\right).
\ee
 In terms of these generators,  all of primitive elements in the Schottky group could be constructed. However, up to $e^{-6\pi TR}$ only $L_j$'s contribute. There are totally $2n$ such kind of generators including $L_j$ and $L_j^{-1}$. For every $L_j$, its larger eigenvalue is
\bea q_{\gamma}^{-\frac{1}{2}}&=&\frac{1}{2}[(L_j)_{11}+(L_j)_{22}+\sqrt{((L_j)_{11}+(L_j)_{22})^2-4}] \notag \\
&=&\frac{nu_y}{(1-u_y^2)\sqrt{u_R}}(u_y^{-\frac{1}{n}}-u_y^{\frac{1}{n}})
\left\{1+\frac{1}{n^2u_y^2}(1-u_y^2)^2u_R(u_y^{-\frac{1}{n}}-u_y^{\frac{1}{n}})^{-2} \right. \notag \\ &~&\left(\frac{n^2u_y^2}{(1-u_y^2)^2}(u_y^{-\frac{1}{n}}-u_y^{\frac{1}{n}})(-u_y^{-\frac{1}{n}}\frac{[(n+1)u_y^2+(n-1)]^2}{4n^2u_y^2} \right.\notag\\
&~&\left.\left.+u_y^\frac{1}{n}\frac{[(n-1)u_y^2+(n+1)]}{4n^2u_y^2}) -1\right)\right\}
\eea
which is independent of $j$.  Then the 1-loop partition function is
\bea \log~Z_n\mid_{1-loop}&=&\sum_{\gamma}\sum_{m=2}^{\inf}(-1)\log\mid 1-q_{\gamma}^m\mid \notag \\
&=&\sum_{\gamma}Re~[q_{\gamma}^2+q_{\gamma}^3]+O(e^{-\frac{8\pi L}{\beta}}) \notag \\
&=&2n\left\{\frac{1}{n^4}(\frac{\sinh2\pi Ty}{\sinh\frac{2\pi Ty}{n}})^4 e^{\frac{-4\pi L}{\beta}}
+\frac{5}{n^6}(\frac{\sinh2\pi Ty}{\sinh\frac{2\pi Ty}{n}})^6e^{\frac{-6\pi L}{\beta}} \right.\notag \\
&~&+\frac{1}{n^4}(\frac{\sinh2\pi Ty}{\sinh\frac{2\pi Ty}{n}})^5
\left(\frac{\sinh(\frac{2\pi Ty}{n})}{\sinh(2\pi Ty)}4\cosh^2(2\pi Ty)
-\frac{8}{n}\cosh(2\pi Ty)\cosh(\frac{2\pi Ty}{n}) \right.\notag \\
&~&\left.\left.+\frac{4}{n^2}\sinh(2\pi Ty)\sinh(\frac{2\pi Ty}{n})\right)e^{-\frac{6\pi L}{\beta}}\right\}+O(e^{-\frac{8\pi L}{\beta}}).
\eea
When n=1
\be \log~Z_1\mid_{1-loop}=2e^{\frac{-4\pi L}{\beta}}+2e^{-\frac{6\pi L}{\beta}}+O(e^{-\frac{8\pi L}{\beta}}). \ee
Thus we have
\bea\label{quan} &~&S_n\mid_{1-loop}=-\frac{1}{n-1}(\log Z_n-n\log Z_1) \notag \\
&~&=-\frac{2n}{n-1}\left(\frac{1}{n^4}\frac{\sinh^4(2\pi Ty)}{\sinh^4\frac{2\pi Ty}{n}}-1\right)e^{-4\pi TR} \notag \\
&~&-\frac{2n}{n-1}\left(\frac{4}{n^4}\frac{\sinh^4({2\pi}Ty)}{\sinh ^4\frac{{2\pi}Ty}{n}}\cosh^2({2\pi}Ty)
-\frac{8}{n^5}\frac{\sinh^5({2\pi}Ty)}{\sinh^5\frac{2\pi Ty}{n}}\cosh(2\pi Ty) ~\cosh\frac{2\pi Ty}{n} \right.\notag \\
&~&\left.+\frac{4}{n^6}\frac{\sinh^6(2\pi Ty)}{\sinh^4\frac{2\pi Ty}{n}}+
\frac{5}{n^6}\frac{\sinh^6({2\pi}Ty)}{\sinh^6\frac{2\pi Ty}{n}}-1\right)e^{-6\pi TR} +O(e^{-8\pi TR}).
\eea
The entangle entropy is
\bea S_{EE}&=&\frac{c}{3}\log\sinh2\pi Ty+const+
8[1-2\pi Ty\coth(2\pi Ty)]e^{-4\pi TR} \notag \\
&~&+12[1-2\pi Ty\coth(2\pi Ty)]e^{-6\pi TR}. \label{EEhigh}
\eea

\subsection{Low temperature Renyi entropy}
 At low temperature, the partition function and the R\'enyi entropy can be obtained by acting the duality transformation (\ref{Sdual}) on the results at high temperature. The transformation leads to
\bea &~&e^{-2\pi TR}\rightarrow e^{-\frac{2\pi}{TL}}, \notag \\
&~&\cosh^2{2\pi}Ty\rightarrow \cos^2\frac{\pi l}{L}, \notag \\
&~&\sinh^2{2\pi}Ty\rightarrow -\sin^2\frac{\pi l}{L}. \eea
Considering these transformations, it is easy to see that the classical part (\ref{class}) is in perfect agreement with (\ref{tree}), and the 1-loop correction (\ref{quan}) is in perfect agreement with (\ref{1loop}). 

\section{Conclusion and discussion}

In this work, we studied the R\'enyi entropy and entanglement entropy of one single interval on a circle at low temperature in 2D CFT. When the temperature is low, we are allowed to expand the thermal density matrix level by level. We focused on the vacuum Verma module and considered the excitations up to level four. Such a consideration was motivated by the holographic computation of R\'enyi entropy in pure AdS$_3$ gravity. We found exact agreement in the large $c$ limit between field theory and holographic computation up to $e^{-\frac{8\pi\beta}{L}}$ for classical contribution  and to $e^{-\frac{6\pi\beta}{L}}$ for 1-loop correction. Our discussion in this work provides another evidence  that three dimensional pure AdS$_3$ gravity correspond to a conformal field theory with only vacuum Verma module\cite{Zhang,Hartman}.

One important ingredient in our holographic computation is to consider the size (temperature) dependence of the R\'enyi entropy. By considering the monodromy condition and the variation with respect to the worldsheet metric, we propose the differential equations (\ref{size}) and (\ref{tem}), which help us to determine the $y$-independent constants in the classical action. Our treatment should apply for other cases at finite temperature and finite size.

The discussion in the present work is technically very different from the ones in \cite{Zhang}.  Considering the boundary Riemann surfaces of the gravitational configurations, the one in this work is to connect $n$ genus-1 torus  along one branch cut; while in the two-interval case\cite{Zhang} we connect $n$ sheets  at two branch cuts. When we calculate $S_n$, even though the replica symmetry is $\mathbb{Z}_n$, the Riemann surface at finite temperature is of genus $n$; while for double intervals it is of genus $n-1$. Moreover the field calculation is also quite different. In this work we did the low temperature expansion and considered the excitations level by level. At each level, the computation boils down to the correlation of vertex operators inserted at the infinity past and infinity future on some sheet.  While in \cite{Zhang} the twist operators being inserted at the branch points in every interval are close to each other so that one may use the operators product expansion of the twist operators to compute the R\'enyi entropy order by order in small cross ratio.

One interesting issue is on the R\'enyi entropy of one single interval on a circle at high temperature. From the holographic point of view, the computation in the high temperature case is dual to the one in the low temperature case by a modular transformation. This is possible because the main difference in the bulk computation comes from the monodromy condition. In the high temperature case, one needs to impose the trivial monodromy along the time circle, while in the low temperature case, it is the trivial monodromy along the space circle. In both cases, one impose the trivial monodromy along the cycle enclosing the branch cut from single interval. As a result, the two gravitational configurations in two cases are dual to each other. However, on the CFT side, all the excitations have to be taken into account at high temperature such that a direct computation seems to be impossible. Naively, one may think that the modular transformation exchanges the role of time and space direction, especially after Euclideanization. For example, it is conceivable that the thermal density matrix could be taken as $\propto e^{-LH}=e^{-2\pi (L/ \b) (L_0+\tilde L_0)}$. But the interval which is along the spatial circle would be mapped to the interval along the time circle. This is exactly the dual geometry showing us. This could be seen from the functional combination of $y$ with $L$ and $T$ in two cases.  However, it is clear in the bulk computation, the interval was treated as in the spatial circle. Nevertheless, considering the accumulating evidence on the correctness of holographic computation of two-interval R\'enyi entropy  studied in \cite{Zhang,Chen:2013dxa,Chen:2014kja} and the single-interval R\'enyi entropy at finite temperature case discussed above, we are inclined to believe that holographic computation of R\'enyi entropy in other cases, including the high temperature case, provides a trustable way to compute exactly the same entropy in 2D CFT in the large central charge limit.
Therefore, the holographic computation  at high temperature in \cite{Dong} allows us to read the thermal corrections to the R\'enyi entropy and the entanglement entropy from (\ref{class},\ref{quan}) and (\ref{EEhigh}) respectively.

One furthermore question is on how to calculate a large interval R\'enyi entropy at high temperature. In \cite{twostr}, the holographic entanglement entropy in this case has been studied. It was found that  the geodesic in the bulk would break into two pieces, one winding  around the black hole, the other  connecting two boundary points. Therefore
$S_{th}=\lim_{\epsilon\rightarrow 0}(S(1-\epsilon)-S(\epsilon)).$
To compute the holographic Renyi entropy in this case, it seems that we need to find another set of monodromy condition to build the bulk structure\cite{ChenWu}. 

In this work, we paid more attention to the excitations of the vacuum Verma module and
checked the AdS/CFT correspondence of pure AdS$_3$ gravity. It would be interesting to generalize the present study to the case with other primary fields\cite{ChenWu}. The similar studies in the two-interval case have been pursued in the context of HS/CFT correspondence\cite{Chen:2013dxa,Perlmutter:2013paa}, the AdS$_3$/LCFT$_2$ correspondence\cite{Chen:2014kja}, and the scalar matter coupling\cite{Beccaria:2014lqa}. The correction from chemical potential to higher spin entanglement entropy in the single interval on the infinite spatial line and at finite temperature has been studied recently in \cite{Datta:2014uxa}.



\newpage

\noindent {\large{\bf Acknowledgments}}\\

We would like to thank Jia-ju Zhang for many valuable discussions.
BC would like to thank the participants of the Workshop on "Holography for black holes and cosmology"(ULB, Brussels) for stimulating discussions. The work was in part supported by NSFC Grant No.~11275010, No.~11335012 and No.~11325522. BC would like to  thank the organizer and participants of the advanced workshop ``Dark Energy and Fundamental Theory" supported by the Special Fund for Theoretical Physics from the National Natural Science Foundations of China with grant no.: 10947203 for stimulating discussions and comments.
\vspace*{5mm}

\begin{appendix}
\section{Vertex operators}

In this appendix, we present some technical details on the vertex operators.

\subsection{Vertex operator at infinity}

 The vertex operators from the vacuum module at origin are well known
\bea\label{vertex0} L_{-2}\mid 0\rangle&\rightarrow& T(u), \notag \\
L_{-3}\mid 0\rangle&\rightarrow& \partial T(u), \notag \\
L_{-4}\mid 0\rangle&\rightarrow& \frac{1}{2}\partial^2 T(u), \notag \\
L_{-2}L_{-2}\mid 0\rangle&\rightarrow& :T(u)^2:. \eea

At infinity, we just need to make a conformal transformation $w=\frac{1}{u}$, and get
\bea\label{vertexinf1} T(u)&\rightarrow& T(w)(\frac{\partial w}{\partial u})^2=w^4T(w), \notag \\
\partial T(u)&\rightarrow& \frac{\partial w}{\partial u}\partial_w(T(w)(\frac{\partial w}{\partial u})^2)
=-w^6\partial T(w)-4w^5T(w),  \\
\frac{1}{2}\partial^2 T(u)&\rightarrow&\frac{\partial w}{\partial u}
\partial_w[\frac{\partial w}{\partial u}\partial_w(T(w)(\frac{\partial w}{\partial u})^2)]
=\frac{1}{2}w^8\partial^2T(w)+5w^7\partial T(w)+10w^6T(w). \notag
\eea
As $w=\frac{1}{u}$ belongs to $SL(2,C)$ transformation, there is no Schwarzian derivative term.

For $:T(u)^2:$, the transformation is a little complicated
\bea\label{vertexinf2} &~&:T(u)^2:=\frac{1}{2\pi i}\oint\frac{1}{u_1-u}du_1T(u_1)T(u) \notag \\
& &=\frac{1}{2\pi i}\oint\frac{\partial u_1}{\partial w_1}dw_1\frac{1}{\frac{1}{w_1}-\frac{1}{w}}
T(w_1)(\frac{\partial w_1}{\partial u_1})^2T(w)(\frac{\partial w}{\partial u})^2 \notag \\
& &=\frac{1}{2\pi i}\oint dw_1(-1)\frac{ww_1}{w_1-w}\frac{\partial w_1}{\partial u_1}(\frac{\partial w}{\partial u})^2
(\frac{c}{2}\frac{1}{(w_1-w)^4}+\frac{2T(w)}{(w_1-w)^2}+\frac{\partial T(w)}{w_1-w}+:T(w_1)T(w):) \notag \\
&&=w^8:T(w)^2+3w^7\partial T(w)+6w^6T(w).
 \eea

\subsection{Conformal transformation for $:T(u)^2:$}

In our computation,  we need the explicit forms of the operators in full complex plane. Under
\be \zeta=(\frac{u-e^{i\theta_2}}{u-e^{i\theta_1}})^{\frac{1}{n}}, \ee
the transformation of the energy-momentum tensor and its higher derivatives are easy to obtain.  However,  we still need to deal with the conformal transformation on $:T(u)^2:$. It changes as
\bea &~&:T(u)^2:\rightarrow\frac{1}{2\pi i}\oint\frac{1}{u_1-u}du_1T(u_1)T(u) \notag \\
&&=\frac{1}{2\pi i}\oint\frac{1}{u_1-u}du_1\left(T(\zeta_1)(\frac{\partial\zeta_1}{\partial u_1})^2
+\frac{c}{24}(1-\frac{1}{n^2})\frac{(e^{i\theta_2}-e^{i\theta_1})^2}{(u_1-e^{i\theta_2})^2(u_1-e^i\theta_1)^2}\right) \notag \\
&~&\times\left(T(\zeta)(\frac{\partial\zeta}{\partial u})^2
+\frac{c}{24}(1-\frac{1}{n^2})\frac{(e^{i\theta_2}-e^{i\theta_1})^2}{(u-e^{i\theta_2})^2(u-e^{i\theta_1})^2}\right) \notag \\
&&=\frac{1}{1440}cA(u)+\left(-\frac{1}{2}(\frac{{\zeta}^{''}}{{\zeta}^{'}})^2+\frac{4}{3}\frac{\zeta^{(3)}}{\zeta^{'}}\right)(\zeta^{'})^2T(\zeta)
+\frac{3}{2}\frac{\zeta^{''}}{\zeta^{'}}(\zeta^{'})^3\partial T(\zeta)+(\zeta^{'})^4:T(\zeta)^2: \notag \\
&~&+2T(\zeta)(\frac{\partial\zeta}{\partial u})^2\frac{c}{24}(1-\frac{1}{n^2})
\frac{(e^{i\theta_2}-e^{i\theta_1})^2}{(u-e^{i\theta_2})^2(u-e^{i\theta_1})^2} \notag \\
&~&+(\frac{c}{24})^2(1-\frac{1}{n^2})^2(\frac{(e^{i\theta_2}-e^{i\theta_1})^2}{(u-e^{i\theta_2})^2(u-e^{i\theta_1})^2})^2 \eea
where
\bea A(u)&=&\frac{1}{n^4}(n^2-1)(\frac{1}{u-e^{i\theta_1}}-\frac{1}{u-e^{i\theta_2}})^2
\left(2(11+61n^2)\frac{1}{(u-e^{i\theta_1})(u-e^{i\theta_2})} \right.\notag \\
&~&\left.+(-11+119n^2)(\frac{1}{(u-e^{i\theta_1})^2}+\frac{1}{(u-e^{i\theta_2})^2})\right)
\eea

\section{Multi-point functions for vertex operators}

Here we list some useful relations in computing the multi-point functions
\be \langle T(u_1)T(u_2)\rangle=\frac{c}{2}\frac{1}{(u_1-u_2)^4}, \ee
\be \langle T(u_1)T(u_2)T(u_3)\rangle=\frac{c}{(u_1-u_2)^2(u_2-u_3)^2(u_3-u_1)^2}, \ee
\bea\label{fourpoint} &~&\langle T(u_1)T(u_2)T(u_3)T(u_4)\rangle \notag \\
&&=\frac{c^2}{4}[\frac{1}{(u_1-u_2)^4(u_3-u_4)^4}+\frac{1}{(u_1-u_3)^4(u_2-u_4)^4}+\frac{1}{(u_1-u_4)^4(u_2-u_3)^4}] \notag \\
&~&+c\frac{(u_2-u_3)^2(u_1-u_4)^2+(u_2-u_4)^2(u_1-u_3)^2+(u_3-u_4)^2(u_1-u_2)^2}
{(u_1-u_2)^2(u_1-u_3)^2(u_1-u_4)^2(u_2-u_3)^2(u_2-u_4)^2(u_3-u_4)^2}, \eea
\bea\label{twopoint1} \langle :T(u_2)^2:T(u_3)\rangle=\frac{1}{2\pi i}\oint \frac{du_1}{u_1-u_2}\langle T(u_1)T(u_2)T(u_3)\rangle=\frac{3c}{(u_2-u_3)^6}, \eea
\bea\label{twopoint2} \langle :T(u_2)^2::T(u_4)^2:\rangle &=&\frac{1}{2\pi i}\oint \frac{du_1}{u_1-u_2}\frac{1}{2\pi i}\oint \frac{du_3}{u_3-u_4}
\langle T(u_1)T(u_2)T(u_3)T(u_4)\rangle \notag \\
&=&(\frac{1}{2}c^2+40c)\frac{1}{(u_2-u_4)^8}. \eea

\section{Summation formulae}

In calculation we need two kinds of summation
\bea &~&\sum_{j=1}^{n-1}\frac{1}{(2sin\frac{\pi j}{n})^4}, \hs{5ex}
\sum_{j=1}^{n-1}\frac{1}{(2sin(\frac{\theta_2-\theta_1}{2n}-\frac{\pi j}{n}))^4}. \eea
By studying the residue of the function
\be f(t)=\frac{n}{t(1-t)^2(1-\frac{1}{t})^2(t^n-1)}, \nn\ee
it is easy to get
\be\label{summation1} \sum_{j=1}^{n-1}\frac{1}{(sin\frac{\pi j}{n})^4}=\frac{1}{720}(n^2+11)(n^2-1). \ee
Similarly,
by studying
\be f(t)=e^{-i(\theta_2-\theta_1)}\frac{n}{t(1-t)^2(1-\frac{1}{t})^2(t^n-e^{-i(\theta_2-\theta_1)})}, \nn \ee
we get
\be\label{summation2} \sum_{j=1}^{n-1}\frac{1}{(2sin(\frac{\theta_2-\theta_1}{2n}-\frac{\pi j}{n}))^4}
=\frac{1}{(2sin\frac{\theta_2-\theta_1}{2})^4}\frac{n^2}{3}[(n^2-1)\cos(\theta_2-\theta_1)+(1+2n^2)]
-\frac{1}{(2sin\frac{\theta_2-\theta_1}{2n})^4}. \ee
\end{appendix}




\vspace*{5mm}
\begin{thebibliography}{99}

\bibitem{nielsen2010quantum}
M.~A. Nielsen and I.~L. Chuang, {\em Quantum computation and quantum
  information}.
\newblock Cambridge university press, 2010.

\bibitem{petz2008quantum}
D.~Petz, {\em Quantum information theory and quantum statistics}.
\newblock Springer, 2008.

\bibitem{Faulkner}
 Thomas Faulkner,
 ``The Entanglement Renyi Entropies of Disjoint Intervals in AdS/CFT''
 [arXiv:1303.7221]

\bibitem{Hartman}
T.~Hartman,
``Entanglement Entropy at Large Central Charge,''
arXiv:1303.6955[hep-th]

\bibitem{Calabrese:2009ez}
P.~Calabrese, J.~Cardy, and E.~Tonni, ``{Entanglement entropy of two disjoint
  intervals in conformal field theory},''
  \href{http://dx.doi.org/10.1088/1742-5468/2009/11/P11001}{{\em J. Stat. Mech.}
  {\bfseries 0911} (2009) P11001},
\href{http://arxiv.org/abs/0905.2069}{[{\ttfamily arXiv:0905.2069 [hep-th]}]}.

\bibitem{Headrick:2010zt}
M.~Headrick, ``{Entanglement Renyi entropies in holographic theories},''
  \href{http://dx.doi.org/10.1103/PhysRevD.82.126010}{{\em Phys. Rev. }
  {\bfseries D82} (2010) 126010},
\href{http://arxiv.org/abs/1006.0047}{[{\ttfamily arXiv:1006.0047 [hep-th]}]}.

\bibitem{Ryu:2006bv}
S.~Ryu and T.~Takayanagi, ``{Holographic derivation of entanglement entropy
  from AdS/CFT},'' \href{http://dx.doi.org/10.1103/PhysRevLett.96.181602}{{\em
  Phys. Rev. Lett.} {\bfseries 96} (2006) 181602},
\href{http://arxiv.org/abs/hep-th/0603001}{[{\ttfamily arXiv:hep-th/0603001
  [hep-th]}]}.

\bibitem{Ryu:2006ef}
S.~Ryu and T.~Takayanagi, ``{Aspects of Holographic Entanglement Entropy},''
  \href{http://dx.doi.org/10.1088/1126-6708/2006/08/045}{{\em JHEP} {\bfseries
  0608} (2006) 045},
\href{http://arxiv.org/abs/hep-th/0605073}{[{\ttfamily arXiv:hep-th/0605073
  [hep-th]}]}.

\bibitem{Dong}
 Taylor Barrella, Xi Dong, Sean A.~Hartnoll and Victoria L.~Martin
 ``Holographic entanglement beyond classical gravity,'' JHEP {\bf 1309}, 109 (2013)
 [arXiv:1306.4682 [hep-th]]

\bibitem{Zhang}
B.~Chen and J.-J. Zhang, ``{On short interval expansion of Rényi entropy},''
  \href{http://dx.doi.org/10.1007/JHEP11(2013)164}{{\em JHEP} {\bfseries 1311}
  (2013) 164},
\href{http://arxiv.org/abs/1309.5453}{[{\ttfamily arXiv:1309.5453 [hep-th]}]}.

\bibitem{Calabrese:2010he}
P.~Calabrese, J.~Cardy, and E.~Tonni, ``{Entanglement entropy of two disjoint
  intervals in conformal field theory II},''
  \href{http://dx.doi.org/10.1088/1742-5468/2011/01/P01021}{{\em J.Stat.Mech.}
  {\bfseries 1101} (2011) P01021},
\href{http://arxiv.org/abs/1011.5482}{[{\ttfamily arXiv:1011.5482 [hep-th]}]}.

\bibitem{Chen:2013dxa}
  B.~Chen, J.~Long and J.~-j.~Zhang,
  ``Holographic Rényi entropy for CFT with W symmetry,''
  JHEP {\bf 1404}, 041 (2014)
  [arXiv:1312.5510 [hep-th]].

\bibitem{Perlmutter:2013paa}
  E.~Perlmutter,
  ``Comments on Renyi entropy in AdS$_3$/CFT$_2$,''
  arXiv:1312.5740 [hep-th].

\bibitem{Chen:2014kja}
  B.~Chen, F.~-y.~Song and J.~-j.~Zhang,
  ``Holographic Renyi entropy in AdS$_3$/LCFT$_2$ correspondence,''
  JHEP {\bf 1403}, 137 (2014)
  [arXiv:1401.0261 [hep-th]].

\bibitem{Beccaria:2014lqa}
  M.~Beccaria and G.~Macorini,
  ``On the next-to-leading holographic entanglement entropy in $AdS_{3}/CFT_{2}$,''
  JHEP {\bf 1404}, 045 (2014)
  [arXiv:1402.0659 [hep-th]].

\bibitem{Datta:2013hba}
  S.~Datta and J.~R.~David,
  ``Rényi entropies of free bosons on the torus and holography,''
  JHEP {\bf 1404}, 081 (2014)
  [arXiv:1311.1218 [hep-th]].

\bibitem{Cardyun}
 John Cardy and Chrisopher P.~Herzog,
 ``Universal Thermal Corrections to Single Interval Entanglement Entropy for Conformal Field Theories'', Phys.\ Rev.\ Lett.\  {\bf 112}, 171603 (2014)
  [arXiv:1403.0578 [hep-th]].


\bibitem{Cardy}
P.~Calabrese and J.~Cardy,
``Entanglement entropy and conformal field theory,''
J.~Phys.~A \textbf{42},504004 (2009)
[arXiv:0905.4013]



\bibitem{Krasnov}
Kirill Krasnov,
``Holography and Riemann surfaces,''
Adv.~Theor.~Math.~Phys. \textbf{4}, 929 (2000)
[hep-th/0005106]

\bibitem{Liou}
P.~G.~Zograf and L.~A.~Takhtadzhyan,
``On the uniformization of Riemann surfaces and the Weil-Petersson metric on Teichmuller and Schottky spaces,''
Math. USSR SB. \textbf{60},297 (1988)


\bibitem{Yin}
X.~Yin,
``Partition Functions of Three-Dimensional Pure Gravity,''
Commun.~Num.~Theor.~Phys.~\textbf{2},~285 (2008)
[arXiv:0710.2129[hep-th]]

\bibitem{Giombi:2008vd}
S.~Giombi, A.~Maloney, and X.~Yin, ``{One-loop Partition Functions of 3D
  Gravity},'' \href{http://dx.doi.org/10.1088/1126-6708/2008/08/007}{{\em JHEP}
  {\bfseries 0808} (2008) 007},
\href{http://arxiv.org/abs/0804.1773}{[{\ttfamily arXiv:0804.1773 [hep-th]}]}.

\bibitem{twostr}
T.~Azeyanagi, T.~Nishioka, and T.~Takayanagi
``Near extremal black hole entropy as entanglement entropy via $AdS_2$/$CFT_1$''
Phys. Rev. D {\bf 77},0604005
[arXiv:0710.2956 [hep-th]].



\bibitem{ChenWu}
B.~Chen and Jie-qiang Wu
``Single interval R\'enyi entropy at high temperature,''
 work in progress.

\bibitem{Datta:2014uxa}
  S.~Datta, J.~R.~David, M.~Ferlaino and S.~P.~Kumar,
  ``A universal correction to higher spin entanglement entropy,''
  arXiv:1405.0015 [hep-th].



\end{thebibliography}
\end{document}